\documentclass[12pt,a4paper]{article}

\usepackage{colordvi}
\usepackage[dvips]{pstcol}
\usepackage{appendix}

\begin{document}

\def\micro{{\tt micrOMEGAs}}
\def\micromegas{{\tt micrOMEGAs}}
\def\ra{\rightarrow}
\def\calchep{{\tt CalcHEP}}
\def\comphep{{\tt CompHEP}}
\def\lanhep{{\tt LanHEP}}
\def\slhaplus{{\tt SLHAplus}}

\def\suspect{{\tt SuSpect}}
\def\mbmb{m_b(m_b)}
\def\mt{m_t}
\def\dMb{\Delta m_b}
\def\dMq{\Delta m_q}
\def\delrho{\Delta\rho}
\def\bsgamma{b\to s\gamma}
\def\bsmu{B_s\to \mu^+\mu^-}
\def\gmuon{(g-2)_\mu}
\def\noi{\noindent}

\begin{flushright}
   \vspace*{-18mm}
   Date: \today
\end{flushright}
\vspace*{2mm}

\begin{center}

{\Large\bf SLHAplus: a library for implementing
extensions of the standard model} \\[8mm]

{\large   G.~B\'elanger$^1$, Neil D. Christensen$^2$,  A.~Pukhov$^3$,  A. Semenov$^4$.}\\[4mm]

{\it 1) LAPTH, Univ. de Savoie, CNRS, B.P.110,  F-74941 Annecy-le-Vieux, France\\
     2) Dept. of Physics, University of Wisconsin, Madison, WI 53706, USA\\
     3) Skobeltsyn Inst. of Nuclear Physics, Moscow State Univ., Moscow 119992, Russia\\
     4) Joint Institute for Nuclear Research (JINR) 141980, Dubna,  Russia\\}
\end{center}

\begin{abstract}
We provide a library to facilitate the implementation of  new models  in codes such as  
matrix element and event generators or codes for computing dark matter observables. 
The library contains a SLHA reader routine as well as diagonalisation routines. 
This library is available in CalcHEP and micrOMEGAs. The implementation of models based
on this library is supported by LanHEP and FeynRules.
\end{abstract}

\section{Introduction}

In the very near future the LHC will have the opportunity to test the TeV scale relevant
to address the outstanding issues in the standard model and its extensions: the symmetry breaking problem and the dark matter problem. 
Furthermore numerous astroparticle searches will help refine the
dark matter properties. Confronting the various extensions of the standard model that have been proposed 
with observations at colliders and astroparticle searches will therefore provide powerful tests of 
the new physics model. Sophisticated tools have been developed to compute the predictions for observables such as particle spectra, 
decay rates or cross sections that are relevant at colliders ( for a repository of available tools see ~\cite{Skands:2005vi}) while other tools  explore the implications of  new physics
models for dark matter observables~\cite{Gondolo:2004sc,Belanger:2006is,Belanger:2010gh,Baer:2002fv,Arbey:2009gu}. The interpretation of a new signature  might require 
extending those tools to adapt them to work with different models.  

Tools such as LanHEP~\cite{Semenov:2008jy}, FeynRules~\cite{Christensen:2008py} and SARAH~\cite{Staub:2009bi} have been developed to facilitate the implementation 
of new models in generic matrix element and event generators such as  Madgraph~\cite{Alwall:2007st}, \comphep~\cite{Pukhov:1999gg,Boos:2004kh},
\calchep~\cite{Pukhov:2004ca}, FeynArts/FormCalc~\cite{Hahn:2000kx,Hahn:2009bf}, Whizard~\cite{Kilian:2007gr} and  Sherpa~\cite{Gleisberg:2008ta}.
They require mainly  the implementation of a new Lagrangian and provide the model file adapted to the chosen generator.
While these tools are very powerful they often leave some tedious and repetitive tasks for the user.
 Here we provide  a library that contains some routines  to facilitate the implementation of new models. 
The library first contains a routine for reading a SUSY Les Houches Accord (SLHA) file~\cite{Skands:2003cj,Allanach:2008qq}. 
An SLHA file is a standardized file format specifying input and output
parameters that was developed in  the framework of supersymmetry to 
facilitate the passing of information beween codes as varied as spectrum calculators, matrix element generators and event generators. 
\footnote{Another SLHA reader library that can be used with the MSSM and its
extensions is available at ~\cite{Hahn:2006nq}.}
We have generalised the procedure to take into account an arbitrary number of blocks so that the reader can be used in generic models
including non supersymmetric ones.
The library also contains routines to diagonalise real and complex mass matrices with either unitary or bi-unitary transformations.
Finally  it contains some routines for evaluating the running strong  coupling constant as well as running quark masses and effective quark masses.
This library was designed and used in
\calchep~\cite{Pukhov:2004ca},  \micro~\cite{Belanger:2006is},  \lanhep~\cite{Semenov:2008jy}
and FeynRules~\cite{Christensen:2008py} but can be incorporated in other tools as well.

In this paper we describe the set of functions used for the SLHA reader 
(section \ref{SLHA}), the QNUMBERS reader (section \ref{QNUMBERS}),  
matrix diagonalisation (section \ref{Matrix}), QCD functions (section \ref{sec:qcd}) and their 
Fortran counterparts (section \ref{Fortran}). In section \ref{Compilation}, we describe how to compile and  link this code.  Section \ref{LanHEP} comments on special features of this package in LanHEP and CalcHEP.  Section \ref{FeynRules} comments on the use of this library when using FeynRules with CalcHEP.  Finally, in section \ref{Conclusion}, we conclude.  We also include an appendix with an example of using this package with LanHEP and CalcHEP for the MSSM.

\section{SLHA reader}
\label{SLHA} 

The SUSY Les Houches Accord ~\cite{Skands:2003cj,Allanach:2008qq} specifies a unique set of conventions for the MSSM and 
its extensions (NMSSM, CPVMSSM) together with generic file structures for
specification of input parameters, supersymmetric particle masses and couplings as well as decay tables. This allows, for example, for 
dedicated external programs to perform an accurate calculation of the particle spectrum.  The results for
particle masses, mixing angles and other model parameters are then written in a file using a standard format that can be used by other codes
~\cite{Djouadi:2002ze,Allanach:2001kg,Porod:2003um,
Paige:2003mg,Ellwanger:2005dv,Lee:2007gn}. 
A similar file format can also be used for other extensions of the MSSM or for non-supersymmetric models.
The routine we describe here allows for reading  of files in  the SLHA format. 

In general a SLHA file contains several pieces of information 
which are called blocks. A block is characterized by its name and if relevant
by an energy scale. Each block contains the values of several physical parameters 
characterized by a {\it key}. The  key consists of a sequence of 
integer numbers. For example for masses the key is the PDG code
~\cite{Eidelman:2004wy}, for mixing matrices the rows and columns of the matrix and for decays the PDG
codes\footnote{The PDG code must be less than 10 digits} of the mother and daughter particles: 
{\small
\begin{verbatim}
BLOCK MASS   # Mass spectrum
#  PDG Code     mass             particle
        25     1.15137179E+02   # lightest neutral scalar
        37     1.48428409E+03   # charged Higgs
  
BLOCK NMIX  # Neutralino Mixing Matrix
  1  1     9.98499129E-01   # Zn11
  1  2    -1.54392008E-02   # Zn12

BLOCK Au Q=  4.42653237E+02  # The trilinear couplings
  1  1    -8.22783075E+02   # A_u(Q) DRbar
  2  2    -8.22783075E+02   # A_c(Q) DRbar
  
DECAY        36     2.10E-06   # Lightest pseudoscalar
     9.54085917E-03    2          21        21   # BR(A_1 -> gluon gluon)
     2.12111874E-04    2          13       -13   # BR(A_1 -> muon muon)
     5.88191379E-02    2          15       -15   # BR(A_1 -> tau tau)
     6.97107293E-04    2           3        -3   # BR(A_1 -> s sbar)
     1.62828588E-03    2           4        -4   # BR(A_1 -> c cbar)
     9.29088642E-01    2           5        -5   # BR(A_1 -> b bbar)  
\end{verbatim}
}

Strictly speaking the SLHA~\cite{Skands:2003cj}  does not specify the structure of 
blocks. \footnote{We thank Peter Skands for pointing that out to us.}  
However in order to write an universal SLHA reader one has to restrict the
structure of blocks.
We impose the additional requirements that the number 
that describe a  physical quantity is the last number for each block record and 
that there is only one such number for each record. 
We also require that all textual information is included as a comment.
All examples of SLHA blocks described in ~\cite{Skands:2003cj,Allanach:2008qq} satisfy these conditions. 
Note however that for decays the SLHA does not follow the normal BLOCK structure, 
for each record the branching ratio is given as the first number. \slhaplus~ can read correctly
the format of the block \verb|DECAY| as displayed in the example above.

Finally we have implemented in the SLHA reader the possibility to work with complex numbers. 
Complex numbers have to be written  in a Fortran-like format, with a comma 
separating the  real and imaginary parts  and placing the components
into brackets.
\verb|(real, imaginary)|.\\

The functions described below allow to read the information contained in this file, including the blocks that contain the
model parameters, the masses and other physical parameters as well as other information such as particle decay widths.

\noindent
$\bullet$ \verb|slhaRead(filename,mode)|\\
reads all or  part of the data  from the file \verb|filename|.
\verb|mode| is an integer which determines which part of the data should be read from the file, 
\verb|mode= 1*m1+2*m2+4*m4+8*m8+16*m16|  where
\begin{verbatim}
  m1 = 0/1 -   overwrites all/keeps old data    
  m2 = 0/1 -   ignore errors in input file/ stop in case of error                                                                            
  m4 = 0/1 -   read DECAY /do not read   DECAY
  m8 = 0/1 -   read BLOCK/do not  read   BLOCK
  m16 = 0/1 -  read QNUMBERS/do not  read   QNUMBERS
\end{verbatim}

For example \verb|mode=20| (\verb|m4=1,m16=1|) is an instruction to overwrite all previous data and 
read only the information stored in the BLOCK sections of
\verb|filename|. In the same manner \verb|mode=25=1+8+16| is an instruction to add information from DECAY 
to the data obtained previously.

 The function \verb|slhaRead| returns the values:
\begin{verbatim}
  0  - successful reading
 -1  - can not open file
 -2  - invalid data as indicated by SPINFO
 -3  - no data
 n>0 - wrong file format at line n
\end{verbatim}
\noindent
$\bullet$ \verb|slhaValExists(BlockName, keyLength, key1, key2,...)|\\
checks the existence of specific data in a given block. 
\verb|BlockName| can  be substituted with any case spelling.
The \verb|keyLength| parameter defines the length of the key set
\verb|{key1,key2,...}|. 
For example
     \verb|slhaValExists("Nmix",2,1,2)|
will return 1 if the neutralino mass mixing element \verb|Zn12| is given in the file and 
0 otherwise.

\noindent
$\bullet$ \verb|cslhaVal(BlockNmae,Q, keyLength, key1, key2,......)|\\
is the main routine which allows to  extract the  numerical values of parameters.
\verb|BlockName| and \verb|keyLength| are defined above.
The parameter \verb|Q| defines  the scale dependence. 
This parameter is relevant only for the blocks that contain scale dependent parameters, it will be ignored for other
blocks, for example those that give the particle pole masses. 
In general a SLHA file can contain several blocks with 
the same name but different scales (the scale is specified after the name of the block).
\verb|slhaVal| uses the following algorithm to read the scale dependent parameters. 
If \verb|Q| is less(greater) than all the  scales used in the different blocks for a given parameter 
\verb|slhaVal| returns the value corresponding to the minimum(maximum) scale contained in the file.
Otherwise \verb|slhaVal| reads the values corresponding to the two scales $Q_1$ and $Q_2$ just below and above
\verb|Q| and performs a linear interpolation in log(\verb|Q|) to evaluate the 
returned values. \verb|cslhaVal| returns complex number.
If the data file contains  real number then  imaginary part of  return
value is zero.

$\bullet$ \verb|slhaVal(BlockNmae,Q, keyLength, key1, key2,......)|\\
This function is defined as  \verb|creal(cslhaVal(...))|. 
 It can be used instead of \verb|cslhaval| for all models
with only real parameters.

\noindent
$\bullet$ \verb|slhaWarnings(FD)|\\                                                                              
writes into the file \verb|FD| \footnote{ FILE* type in C and channel number in
Fortran} the warnings or error message  stored in the SPINFO block and
returns the number of warnings. If \verb|FD=NULL| the warnings are not written in a file.

\noindent
$\bullet$ \verb|slhaWrite(Filename)|\\
 writes down the information stored by 
\verb|readSLHA| into the file. This function can be used for testing purposes.

\noindent
$\bullet$ \verb|slhaDecayExists(pNum)|\\
checks whether information about the  decay of particle \verb|pNum| exists in the SLHA file. 
\verb|pNum| is the particle PDG code. This function returns the number 
of decay channels listed. 
In particular zero means that the SLHA file contains  information only about the total width, not on branching ratios while 
-1 means that even the total width is not given.

\noindent
$\bullet$ \verb|slhaWidth(pNum)|\\ returns the value of the particle width.

\noindent
$\bullet$ \verb|slhaBranch(pNum,N, nCh)|\\
returns the  branching ratio of  particle \verb|pNum| into the N-th decay channel. Here\\
\noindent \verb|0<N<=slhaDecayExists(pNum)|.
The array \verb|nCh| is an output which specifies the  PDG numbers of the decay products, the list  
is terminated by zero.

The functions \verb|slhaValExists|, \verb|slhaVal|, 
\verb|slhaDecayExists|, \verb|slhaWidth| can be used directly 
in \calchep~ model files.
For example, the mass of the lightest neutralino can be specified in the \calchep\ file \verb|func1.mdl| as

\begin{verbatim}
 MNE1  |  slhaVal("MASS",QSUSY,1,1000022)
\end{verbatim}

Some applications  might  need the full list  
of SLHA blocks and DECAY items obtained with \verb|slhaRead|. 
The following functions serve this purpose\\
$\bullet$\verb| allBlocks(K,L, blockName, &keyLength, keyArray, &val)|\\
$\bullet$\verb| allDecays(K,L, &pdg, &decayLength,&decayArray,&width,&branching)|\\
where $K$ specifies the block or decay to be read in numerical order.  A $K$ of 1 specifies that the first block (or decay) should be read, a $K$ of 2 specifies that the second should be read, and so on.  $L$ specifies which record inside the block (or decay) should be read, again in numerical order.  These functions return $1$ as long
as  the requested  information exists and return  $0$ otherwise. For the blocks, the name of the block is stored in the string \verb|blockName|, the number of keys is recorded in \verb|keyLength|, the array of keys is stored in \verb|keyArray| and \verb|val| contains the value.  For decays, \verb|pdg| stores the pdg of the particle, \verb|decayLength| specifies the number of outgoing particles, \verb|decayArray| the array of outgoing
particles, and \verb|width| and \verb|branching| the total width and branching for each channel.
If $L=0$, the function \verb|allBlocks| fills only the parameters 
\verb|blockName| and \verb|val|. In this case \verb|val| corresponds to the 
scale at which the data in the block are specified.  When $L=0$, \verb|allDecays| 
only fills the \verb|pdg| and \verb|width| parameters. 

In general, SLHA files also contain textual comments. 
After a call of \verb|allBlocks| or \verb|allDecays|
this  information 
is stored in the global variable \verb|slhaComment|. 
In the Fortran version, this variable is contained in \verb|COMMON/SLHACOMMENT/|.
In particular, for $L=0$ \verb|allBlocks/allDecays| will return the general comment from the file,
while if $L>0$, these functions will return the comment for the $L^{th}$ record.

\subsection{Writing an input SLHA file}

This package contains  three routines which allow to write an SLHA input file 
and launch a spectrum calculator. 

\noindent
$\bullet$ \verb|openAppend(fileName)|\\
deletes the input file \verb|fileName| if it exists 
and creates a new empty file with the same name.
The string  {\it fileName} is stored in memory for subsequent usage with the function \verb|aPrintF|. 

\noindent
$\bullet$ \verb|aPrintF(format,...)|\\
opens the file \verb|fileName| and writes at the end of the file the input parameters needed in the SLHA format or in any other format
understood by the spectrum calculator.  The arguments of 
\verb|aPrintF| are similar to the arguments of the standard \verb|printf| function.

\noindent
$\bullet$ \verb|System(format,...)| \\
generates a shell command using \verb|format| and subsequent arguments.
This command  is then launched by the standard \verb|system| C-function. 

For example, to write directly the SLHA model file needed by \suspect~\cite{Djouadi:2002ze} to compute 
the spectrum in a CMSSM(SUGRA) model, one needs to specify the standard model input parameters ($m_b(m_b), m_t(pole)$) as well as the SUSY input
parameters $m_0,m_{1/2},\tan\beta, {\rm sign}(\mu)$, $A_0$. For this one must
 add the following sequence in the \calchep~ \verb|func1.mdl| model file.

\begin{verbatim}
open  |openAppend("suspect2_lha.in")                                                                                                                                                                                                                             |
input1|aPrintF("Block MODSEL  # Select model\n  1  1   # SUGRA\n")                                                                                                                                                                              |
input2|aPrintF("Block SMINPUTS\n 5 %E#mb(mb)\n 6 %E#mt(pole)\n",MbMb,Mtp)                                                                                                                                                               |
input3|aPrintF("BLOCK MINPAR\n 1 %E #m0\n 2 %E #m1/2\n ",Mzero,Mhalf)                                                                                                                                          |
input4|aPrintF("3 %E #tb\n 4 %E #sign(mu)\n 5 %E #A0\n",tb,sgn,A0)                                                                                                                                          |
sys   |System("%s/suspect2.exe",path())                                                                                                                                                                                                                                 |
rd    |slhaRead("suspect2_lha.out",0)                                                                                                                                                                                                                           |
\end{verbatim}

In this example \verb|path()| specifies the path to the directory where the \suspect~ executable is
located.

\section{ SLHAplus for parton shower generators.}
\label{QNUMBERS}
    
Information about parton level  events 
can be transferred to parton shower generators via a file written in  the XML format~\cite{Alwall:2006yp}.
In a generic extension of the Standard Model,  the parton
shower generator also needs information about the quantum numbers  of new particles, 
their  masses, widths and decay modes. 
In Ref.~\cite{Alwall:2007mw} it was proposed  to 
include the corresponding SLHA {\tt Block} and {\tt Decay} items in the 
header section of the XML event file, using the subtags
\verb|<slha>| and \verb|</slha>| to delimit the beginning and the end of the corresponding subsection of the header. 
A new SLHA block, \verb|QNUMBERS|, was designed to provide information about  the
quantum numbers  of new particles. For example a new neutral scalar particle called \verb|balleron| could be
defined as
\begin{verbatim}
BLOCK QNUMBERS 7654321 # balleron
      1     0 # 3 times electric charge
      2     1 # number of spin states (2S+1)
      3     1 # colour rep (1: singlet, 3: triplet, 8: octet)
      4     0 # Particle/Antiparticle distinction (0=own anti)
\end{verbatim}
Strictly speaking the block \verb|QNUMBERS| does not fit the  definition of a standard  
SLHA block as used in this paper.  We treat it as an exceptional block.
This block can be read by  the \verb|slhaRead| routine  but for this special routines are required, which we now describe.\\

\noindent
$\bullet$ \verb|findQnumbers(PDG,&eQ3,&spinDim,&cDim,&anti)|\\
gives the  quantum numbers \verb|eQ3,spinDim,cDim,anti| for the particle specified by 
its {\tt PDG} code    
(see the example of the \verb|QNUMBERS| block 
given above). \verb|findQnumbers| returns 1 if the file contains the quantum 
numbers for a given {\tt PDG} particle, -1 if instead the information about the antiparticle 
is contained in the file, and 0 if the file contains no data on this (anti)particle. 

A parton shower generator may also need the complete list of new
particles. For this, one can use \\
\noindent 
$\bullet$ \verb|allQnumbers(K,&PDG,&eQ3,&spinDim,&cDim,&anti)|\\
where \verb|K| specifies the \verb|QNUMBERS| block to be read in order.  
If $K=1$, the first \verb|QNUMBERS| block is read, if $K=2$, then the second and so on.
As long as \verb|K| is  less than the number of blocks, the function returns 1, otherwise the function 
returns zero. Note that here the \verb|PDG| code is a return parameter.
If in the SLHA file the \verb|QNUMBER| block contains a comment, it is  stored in
the global variable \verb|slhaComment|.

One can not use {\tt slhaRead} to extract SLHA information from an XML event file since
the event information is written in a different format.  
For this reason we  include another file reader which works
with an already open file until  the tag signalling the
end of the SLHA section in the event file  (\verb|</slha>|) is reached\\
\noindent
$\bullet$ \verb|slhaReadStream(FD,mode,"</slha>")|\\
\verb|FD| is a file descriptor, and \verb|mode| is an integer that determines which data has to be read from the file,
it has the same meaning as in \verb|slhaRead|. This function returns the same value as
\verb|slhaRead|.

\section{Matrix diagonalisation}
\label{Matrix}

In a new model one  often has to diagonalise  mass matrices. These squared matrices can be real or complex 
and their properties depend on whether one deals with fermionic or bosonic fields
In a Lagrangian, mass terms
for $d$ real boson fields $\Phi=(\phi_1,\phi_2...\phi_d)$ are written  as
$$\frac{1}{2}\Phi^T M_{rs} \Phi$$  where $M_{rs}$ is a  real symmetric  matrix.
To get  the physical fields which are mass 
eigenstates  of the Hamiltonian one has to perform  an  orthogonal rotation 
$$ \Phi_{phys}= V\Phi$$  
such that  
\begin{equation}
\label{pr_rs}
      V M_{rs} V^T =  M_{diag} \\
\end{equation}
where the diagonal mass  matrix has elements 
\begin{equation}
\left(M_{diag}\right)_{\alpha\beta}=  m_\alpha \delta_{\alpha\beta}
\end{equation}
(Here $m_\alpha$ stands for a mass squared.)
For complex boson fields  the  mass matrix is hermitian and the mass term reads
$$  \Phi^{\dagger} M_{h} \Phi$$
In this case the physical fields are 
obtained by an unitary rotation $V$ such that
\begin{equation}
\label{pr_h}
         V M_{h} V^\dagger       = M_{diag} \\
\end{equation}

For spinor fields, and in particular for Dirac fermions, the mass term  has the generic form
$$     {\bar \Psi^R} M_c\Psi^L \;\;\; + \; h.c   $$
where $\Psi^L$ and $\Psi^R$ are the left and right components of the Dirac spinors 
and $M_c$ is a complex matrix.  The 
left and right components of Dirac fields can be rotated independently since  
the kinetic  terms in the Lagrangian are split between the left and right parts.
$$ \Psi^L_{phys} = V \psi^L \;\;\;\;  \Psi^R_{phys}=U \psi^R $$ 
Diagonalisation requires two unitary matrices $U$ and $V$,
\begin{equation}
\label{pr_ca}
          U M_{c} V^\dagger = M_{diag}  \\ 
\end{equation}

In the special case where the mass matrix is real, for example for the chargino sector of the MSSM,
the  $U$ and $V$ matrices are orthogonal and the
diagonalisation condition  reads 
\begin{equation}
\label{pr_ra}
     U M_{r} V^T =  M_{diag}
\end{equation}

Finally in the case of Majorana fermions, the mass matrix  is complex and symmetric, $M_{cs}$.  
Because Majorana fermions are real we have the constraint
$U = V^*$ therefore  only one unitary  matrix is necessary to perform the diagonalisation, 
\begin{equation}
\label{pr_cs}
     V^* M_{cs} V^\dagger  =   M_{diag}.
\end{equation}

\subsection{ Jacobi algorithm and matrix diagonalisation. }

The Jacobi diagonalisation procedure  consists in a sequence of rotations
in 2-dimensional planes each rotation eliminating one  off-diagonal element.
The convergence of the Jacobi method can be simply understood. It relies on the fact that
the sum of the squared elements of a matrix remains the same after 
multiplication by an orthogonal/unitary matrix. Each rotation which eliminates one off-diagonal
element while leaving the others unchanged will therefore decrease the sum of the square of all
off-diagonal elements. 
The Jacobi method is not the  most efficient one for  matrices of high dimensions. 
However in particle physics the dimension of field multiplets is usually not large and  
the  time needed for matrix diagonalisations  is negligible as compared to the time needed for  
calculation of matrix elements and Monte Carlo phase space integration.  

In our library we have included  routines that can perform the diagonalisation of the matrices 
in Eqs.~\ref{pr_rs},\ref{pr_h}, \ref{pr_ca}, \ref{pr_cs}, \ref{pr_ra}, they are respectively 

\noindent
$\bullet$ \verb|rJacobi(d,Mrs,Mdiag,Vo)|\\
$\bullet$ \verb|cJacobiH(d,Mh,Mdiag,Vu)|\\
$\bullet$ \verb|cJacobiA(d,Mc,Mdiag,Uu,Vu)|\\
$\bullet$ \verb|rJacobiA(d,Mr,Mdiag,Uo,Vo)|\\
$\bullet$ \verb|cJacobiS(d,Mcs,Mdiag,Vu)|\\
Here \verb|d| is the  matrix dimension, the second argument is the initial mass matrix. The functions return
\verb|Mdiag| the array of eigenvalues obtained after the diagonalisation and \verb|V| and \verb|U| are
the rotation matrices. The  indices $u,o$ are used to distinguish unitary and orthogonal matrices. 
All matrices are expressed via one-dimensional arrays. The conversion is done with the formula
\begin{equation}
   A_{ij} \to A[i\cdot d + j] \;\;\;\;\; where \;\;\;\;  0 \le i,j < d 
\end{equation}

For the symmetric or hermitian mass matrices ($M_{rs}, M_{cs}, M_h$) only half the off-diagonal elements are independent and need
to be specified, the
conversion to  one-dimensional arrays uses the  formula
\begin{equation}
\label{order_s}
  A_{ij} \to A[i(d-\frac{(i+1)}{2}) +j ]   \;\;\;\;\; where \;\;\;\;  0 \le
i \le j < d
\end{equation}

The $M_h,M_{cs},M_c,V_u,U_u$  matrices have complex elements.
Complex numbers in our package are characterized by the standard C99 type \verb|double complex|. 
The Jacobi functions return zero after a successful diagonalisation and
$1$ otherwise. The  eigenvalues are sorted in increasing order  of their  absolute value.
For (\ref{pr_ra}, \ref{pr_ca}), there is an ambiguity in the phase of the 
eigenvalues. We assume that all  eigenvalues are real and positive.

Our code is based on the \verb|jacobi| routine provided in 
\cite{Numerical}. For another realisation of 
diagonalisation routines for high energy physics see for instance  
\cite{Hahn:2006hr}

\subsection{SLHAplus format for diagonalisation routines.}
\label{slhaDg}
Some packages for matrix element  calculation do not allow  
the use of arrays. Only simple
expressions are allowed. We therefore write the diagonalisation routines presented 
above in a format  readable by MC calculators.

To use the same routine for a matrix of arbitrary size, we 
use a C language option that allows to write routines with an arbitrary number of argument. 
Any diagonalisation routine returns a number which specifies an identifier (ID number) for the  
rotation  and mass matrices obtained in the diagonalisation process.

\noindent 
$\bullet$ \verb|initDiagonal()|\\
 should be called once  before the other 
diagonalisation routines described below. \verb|initDiagonal()| assigns a zero value 
to the internal counter of  eigenvalues and rotation matrices and returns zero.

\noindent
$\bullet$ \verb|rDiagonal(d,M11,M12,..M1d,M22,M23,...Mdd)|\\
$\bullet$ \verb|cDiagonalH(d,M11,M12,..M1d,M22,M23,...Mdd)|\\
$\bullet$ \verb|cDiagonalS(d,M11,M12,..M1d,M22,M23,...Mdd)|\\
diagonalise symmetric(Eq.~\ref{pr_rs}), hermitian (Eq.~\ref{pr_h}) and complex symmetric(Eq.~\ref{pr_cs}) matrices 
of dimension \verb|d| respectively. The  
$d(d+1)/2$  matrix elements, \verb|Mij| $(i\le j)$, are given as arguments.
The functions return an integer number \verb|id| which serves as an  identifier 
for the  eigenvalues vector and rotation matrices. 

\noindent
$\bullet$ \verb|cDiagonalA(d,M11,M12,..M1d,M21,M22,...Mdd)|\\
$\bullet$ \verb|rDiagonalA(d,M11,M12,..M1d,M21,M22,...Mdd)|\\
diagonalise complex (Eq.~\ref{pr_ca}) and real (Eq.~\ref{pr_ra}) non-Hermitian matrices.
Here  all $d^2$ matrix elements are given as arguments. 

For these five different routines the eigenvalues can be obtained by the same function \\
\noindent
$\bullet$ \verb|MassArray(id, i)|  
where $id$ is the identifier associated with the diagonalisation procedure corresponding to any of the
\verb|[cr]Diagonal[HAS]| routines. The 
index $i$ starts with 1. The elements of the  rotation matrices are obtained using the functions 
\\
$\bullet$ \verb|MixMatrix(id,i,j)| - for orthogonal matrices\\
$\bullet$ \verb|MixMatrixU(id,i,j)| - for the \verb|U| orthogonal matrix in Eq.~\ref{pr_ra} \\
$\bullet$ \verb|cMixMatrix(id,i,j)| - for unitary  matrices\\
$\bullet$ \verb|cMixMatrixU(id,i,j)| - for the \verb|U| hermitian  matrix in Eq.~\ref{pr_ca}\\

\subsection{Errors}

The global variable \verb|FError| signals fatal problems that occur in the execution of
\verb|slhaRead|, \verb|slhaVal|, \verb|slhaWrite|, \verb|slhaBranch|, \verb|System|
as well as \verb|[cr]Diagonal[HAS]| and \verb|[c]MixMatrix[U]|. When \verb|FError=1|, numerical calculations in \calchep~ will be interrupted automatically.

\section{QCD functions}
\label{sec:qcd}

Here we describe some QCD functions which can be useful for implementing a new
model.

\noindent$\bullet$ \verb|initQCD(alfsMZ,McMc,MbMb,Mtp)|\\
This function initializes the parameters needed for the functions
listed below. It has to be called before any of these functions.
The input parameters are the QCD coupling at the Z scale,
$\alpha_s(M_Z)$, and the running quark masses, $m_c(m_c), m_b(m_b)$ and
$m_t(pole)$.

\noindent$\bullet$ \verb| alphaQCD(Q)|\\
calculates the  running $\alpha_s$ at the scale \verb|Q| in the
$\overline{MS}$ scheme. The calculation is done using the
\verb|NNLO| formula in~\cite{Hagiwara:2002fs}. Thresholds for
b-quark and t-quark  are included in  $n_f$ at the scales $\mbmb$
and $\mt(\mt)$ respectively.

\noindent$\bullet$ \verb| MtRun(Q), MbRun(Q), McRun(Q) | \\
calculates top, bottom and charm quarks running masses evaluated
at NNLO~\cite{Vermaseren:1997fq}.

\noindent$\bullet$ \verb| MtEff(Q), MbEff(Q), McEff(Q),  | \\
calculates effective top, bottom and charm quark masses using
~\cite{Spira:1997dg}
\begin{eqnarray}
\label{meff}
 M_{eff}^2(Q)&=&M(Q)^2\left[1+5.67a + (35.94-1.36n_f)a^2 \right.\nonumber\\
 &+& \left.(164.14-n_f(25.77-0.259n_f))a^3\right]
\end{eqnarray}
where $a=\alpha_s(Q)/\pi$,    $M(Q)$  and $\alpha_s(Q)$    are the
running quark masses and  strong coupling  in the
$\overline{MS}$-scheme. These effective masses at the scale $Q=M_h$ give the Yukawa couplings
that reproduce the QCD corrections to the  partial decay width of the Higgs.

\section{Features of the C, C++ and Fortran versions}
\label{Fortran}
All routines of the \slhaplus~ library can be used either in C, C++ or  Fortran.
In general the C,C++ and Fortran routines have the same names
except for the functions with a varying number of arguments which are not supported in Fortran.
Furthermore C++ cannot handled such functions if a parameter is a complex number. 

For each function with a varying number of parameters we include in our
package a set of functions with a fixed number of parameters. The  parameter which
specifies the number of arguments  is removed from the list of arguments
and is instead attached to the function name. For example the generic function 
\verb|slhaVal(BlockName, Q, keyLength,...)| has to   be replaced by\\
\noindent
\begin{verbatim}
slhaVal0(Blockname,Q)
......
slhaVal3(Blockname,Q,key1,key2,key3)
\end{verbatim}
\noindent
For  \verb|slhaVal| and \verb|slhaValExists| the digit attached to the name
specifies the \verb|keyLength|  and can be 0, 1 ,2 or 3. 
For the functions  \verb|[cr]Diagonal[ASH]| this number specifies
the dimension of the matrices. 
For the function \verb|aPrintf(format,...)|  the  digit attached describes the number of
arguments following the parameter \verb|format|. In this case all these arguments
have to be of real type\footnote{In Fortran they have to be of  REAL*8
type.}
 
The function \verb|System| is replaced by two functions  
\verb|System1(format)| and \verb|System2(format,txt)| with text parameters.
The second function can be used to provide the destination of the executable.

Another difference between C,C++ and Fortran concerns error messages.
The variable \verb|FError| gets a non-zero value in case of
a fatal error.
In C or C++ \verb|FError| is a global variable  while in Fortran it is included as 
\begin{verbatim}
integer FError
COMMON/FError/Ferror
\end{verbatim}

The file \verb|SLHAplus.h| contains the function prototypes for C routines.
It has to be included in C and C++ routines which use \slhaplus.
The file \verb|SLHAplus.fh| includes a \verb|Fortran| description of the types
of Fortran functions and \verb|COMMON/FError/| declaration.
It has to be included in each Fortran routines.

\section{Compilation and testing}
\label{Compilation}

The command \verb|make| generates the static library file {\tt libSLHAplus.a} which contains all the
functions of the package.

The subdirectory \verb|test| contains three test programs each available as C, C++ or Fortran  codes.

$\bullet$ \verb|tJacobi.[c/cpp/F]|\\
tests all Jacobi  functions included 
in the package. Used in a cycle it fills the corresponding matrices by  random numbers, 
diagonalises them, prints eigenvalues, restores the non-diagonal matrix  and 
calculates the difference between the original ($M$)  and restored ($M'$)
matrices, $\Delta=\sqrt{\sum_{i,j} \left(
M_{ij}-M'_{ij}\right)^2}$.
A typical  difference  $\Delta \approx 10^{-15}$ means that 
the precision in the diagonalisation reaches the precision of the computer. The dimension of the matrix  and the number of steps 
in the cycle   can be changed  via a redefinition of the parameters {\tt DIM}
and {\tt nTest} which are defined at the top of the file.

$\bullet$ \verb|tChDiag.[c/cpp/F]|\\
 tests the  diagonalisation functions \verb|rDiagonal[A]| of Section 
~\ref{slhaDg}. Here the matrix dimension is 2. This is for a quick test
that the elements of the original matrix have been passed correctly. 

$\bullet$ \verb|tSlha.[c/cpp/F]|\\
 is an independent program for testing the SLHA reader.
The program reads the files {\tt spectr.slha} and {\tt decay.slha} 
and writes on screen the information stored in the files.

To compile  a test program use the command\\
\verb| make main=<name of source code>|\\
The name of the  executable  corresponds to the name of the source file.

\section{Special features of \lanhep~ and \calchep}
\label{LanHEP}

Here we describe some special features relevant for the implementation of a new model in \lanhep~ and \calchep.

\subsection{Particle widths}

The particle widths do not generally need to be defined as input parameters,
indeed the widths depend on the fundamental parameters of the model and 
therefore can be computed automatically by \calchep~ using the relevant Feynman diagrams.
In order to inform \calchep~ that a particle width has to be computed  
before the calculation of a scattering process, the width 
of the particle must be precede by an exclamation mark in the particle table. 
This can be done in \lanhep~ by adding 
the {\tt auto} record after the width declaration.
For instance, 

\noindent
\verb|scalar h/h:  (' Higgs 1',pdg 25,mass Mh,width wh=auto).|

\noindent
will generate the following line in the \verb|prtcls1.mdl| model file.
\begin{verbatim}
Full   Name     | P | aP|  number  |spin2|mass|width|color|
 Higgs 1        |h  |h  |25        |0    |Mh  |!wh  |1    |  
\end{verbatim}

If an SLHA file contains information about particle widths, then the widths can also be defined using the function
\verb|slhaWidth(PDGcode)|.

\subsection{Matrix diagonalisation routines}

As an example we give here the \lanhep~ code for describing the neutralino matrix in the MSSM, the mass eigenvalues 
and the elements of the mixing matrices. 

\begin{verbatim}
parameter NeDiag= rDiagonal(4, MG1,zero, -MZ*SW*cb,  MZ*SW*sb,
                                   MG2, MZ*CW*cb, -MZ*CW*sb,zero, -mu,zero).

_i=1-4 in parameter   MNE_i=MassArray(NeDiag, _i).
	    
_i=1-4,_j=1-4 in parameter  Zn_i_j=MixMatrix(NeDiag, _i, _j).
	   
\end{verbatim}

\noindent
The elements of the mass matrix can either be written explicitly by the user or extracted from \lanhep.
Indeed \lanhep~ contains a facility to examine 
the mass sector of the Lagrangian. When the {\tt CheckMasses } statement is used, 
\lanhep~ creates the file named {\tt masses.chk}. This file contains  
information about mass terms which are not given in the  output files specifying only  the Feynman rules.  
In particular this can be useful for fields that are rotated by orthogonal matrices that diagonalise mass terms, 
\lanhep~ can recognize the mixing matrix if the elements of this matrix were
used in an {\tt OrthMatrix} statement. \lanhep~ restores and prints the 
original mass matrix before the fields rotation. 
For example, in the case of the neutralino this output reads:

\begin{verbatim}
Looking for the mixing matrix...
It is recognized that these fields are rotated by matrix:

  (  Zn11  Zn12  Zn13  Zn14  )
  (  Zn21  Zn22  Zn23  Zn24  )
  (  Zn31  Zn32  Zn33  Zn34  )
  (  Zn41  Zn42  Zn43  Zn44  )

The mass matrix before introducing this rotation:

M11 = (+1*MG1)
M12 = 0
M13 = (-1/CW*MW*SW*cb)
M14 = (+1/CW*MW*SW*sb)
M22 = (+1*MG2)
M23 = (+1*MW*cb)
M24 = (-1*MW*sb)
M33 = 0
M34 = (-1*mu)
M44 = 0
\end{verbatim}

This information can be copied and pasted as argument of the \verb|rDiagonal| function.

\subsection{ External function  prototyping  and linking }

In general one has to provide information about the types 
of parameters and returned values of external functions  to C/Fortran
compilers.  As well   \lanhep~ needs  these prototypes  for numerical 
processing. The functions included in the \slhaplus~ library are {\tt known} 
to \lanhep, \calchep~ and \micromegas, therefore there is no need to specify the prototypes.
The SLHAplus library is included into these packages 
and  linked automatically. If one wants to use  other  external 
functions to calculate couplings or masses the 
following rules have to be used. 

  In \calchep~  the {\tt extlib}N{\tt.mdl} model file ({\it N} is
an integer which specifies the model) can include 
prototypes of C-routines used in the model.\footnote{Note that \lanhep~  does not generate the {\tt
extlib} file and that \calchep~ generates automatically an empty file. }
 Funtion declaration has to start on a new line, the first word  
 has to be {\tt extern} and declaration should be terminated  on 
the same line with a semicolon. For all other details we follow the C-syntax.
Prototypes for functions  which return values of type {\tt double} and have 
all arguments of type {\tt double} can be omitted.
Other records in the  {\tt extlib}N{\tt .mdl}  files are treated as definition 
of external libraries that are provided to the linker.
\micro~ ignores \calchep~  linker instructions and  assumes that all
external functions are stored in the file  {\tt lib/aLib.a}.

The current version of \lanhep~ allows only functions  which return 
{\tt double} and have all arguments of type {\tt double}. Such functions
    have to be declared in \lanhep~ by instructions such as 

\noindent
    \verb|external_func(|{\t name, N, lib}\verb|)| whose arguments describe the
function name, the number of arguments and the name of the shared library which 
contains this function.  In general, the library will be linked when  
processing \lanhep~ for numerical checks. The {\tt lib} parameter 
can be omitted. In this case only symbolic  \lanhep~ processing will be valid.

There are some specific features for writing an external functions in
\lanhep~. First, to pass a string  parameter one has to use the \verb|str| function. For
example to read the SLHA value for one element of the neutralino mixing matrix at the scale \verb|QSUSY| \\
\verb|Zn12 = slhaVal(str(NMIX),QSUSY,2,1,2).|\\
Functions with zero number of parameters should appear in \lanhep~ source files 
without brackets, for example\\
\begin{verbatim}
   external_func(initDiagonal,0).
   parameter zero=initDiagonal.
\end{verbatim}

\subsection{ Complex numbers in \calchep}
Despite the fact that \calchep~ cannot work directly with complex parameters, 
following the procedure described above it is possible to  declare functions with complex parameters and/or 
a complex return value. Complex parameters have to be  constructed   by writing explicitly the  
real and imaginary parts using the parameter $I$ where $I^2=-1$.
Complex return values can be transformed to real values using the  C99  
functions \verb|creal| and \verb|cimag|.  
 Prototypes of these functions are automatically declared in CalcHEP. 
 This way the complex rotation matrices mentionned in section~\ref{slhaDg} can be implemented in CalcHEP.

\section{FeynRules Support}
\label{FeynRules}
Support for the SLHA parameter reading routines of this library has been included in the FeynRules (FR)CalcHEP (CH) output interface which will be released later this year.  Since FR has been described in \cite{Christensen:2008py,Christensen:2009jx}, we will just describe the support for the SLHA parameter files.  The author of a FR model file does not need to do anything different to use this functionality.  It is all handled automatically by the FR CH interface.  There is a new option for the FR CH interface \verb|LHASupport|.  For example, the user could issue the command:
\begin{verbatim}
WriteCHOutput[L,LHASupport -> True]
\end{verbatim}
in their FR session, where \verb|L| is the Lagrangian.  If this option is set to True, the FR-CH interface will write the 
CH model files which use the functions of the SLHAplus library.   In addition, a SLHA parameter file is written 
to the model directory called``varsN.lha" where N is the model number.  To use this model, the user can import 
this model into CH just as with any other CH model.  Since the name of the SLHA parameter file is called explicitly in the model files, 
if the user changes the name of the SLHA file, the user will need to make the corresponding change to the model files.  
In particular, in the file funcN.mdl, the line containing \verb|readSLHA| must be changed to contain the correct filename.
This SLHA parameter file will then need to be copied to the results directory of a CH numerical session.

The SLHA parameter file that FeynRules creates does not have any scales specified for the blocks.  
For this reason, the CH model files that FR writes call the SLHA reader routines with a default scale of 
$0$ (for example, \verb|slhaVal("SMINPUTS",0,1,1)| in the FeynRules Standard Model implementation).  
Since the default parameter file does not specify any scales, this is ignored.  However, 
if a SLHA parameter file from another source was used instead that had blocks with more than 1 scale specified, 
the model files would need to be modified as appropriate.

The matrix diagonalization and QCD routines in \slhaplus~ are not supported in FR at this time.

\section{Conclusion}
\label{Conclusion}

The library \slhaplus~ provides a set of functions to facilitate the implementation of new
models in  codes for computing physical observables. It was developed primarily to be used in
\lanhep~, \calchep~ and \micromegas~ although the routines are general enough to be used with other
codes.  
In the library we also provide the \lanhep~ source code for two different implementation of the
MSSM, these can serve as a basis for the implementation of extensions of the minimal 
supersymmetric model.  The library is available at \verb|lappweb.in2p3.fr/lapth/micromegas/slhaplus|.

\section*{Acknowledgements}
This work was  supported in part by the GDRI-ACPP of CNRS and by the ANR project {\tt ToolsDMColl}, BLAN07-2-194882.
This work  was  also supported by the Russian foundation for Basic Research, grants
RFBR-08-02-00856-a,  RFBR-08-02-92499-a, RPBR-10-02-01443-a and by a State contract No.02.740.11.0244. 
N.D.C was supported by the US NSF under grant number PHY-0705682.

\appendix

\appendixpage
\section{The MSSM model files in LanHEP}

As an example of an application of the routines of the \slhaplus~ library,
we present the \lanhep~ code  to generate the MSSM model files in the format 
of \calchep~ and \micro. The  Lagrangian of the MSSM is described in   
~\cite{Rosiek:1995kg} and some technical details for the implementation into \lanhep~
are  explained in  Ref.~\cite{Belyaev:1997jn, Semenov:2002xn}. 
In this example we describe an MSSM model which takes into account higher order corrections in the Higgs
sector as described below. However we do not include other higher-order corrections such as the 
SUSY-QCD corrections to the $Hb\bar{b}$ vertices~\cite{Guasch:2003cv} or corrections to stop/sbottom interactions. 
Finally we assume massless fermions for the first two generations. 
Therefore the model generated cannot be used for the calculation of the dark matter 
direct detection rates in \micromegas~ which needs non-zero  light quark masses. 
Note that when using the SLHA interface with a spectrum calculator that
includes radiative corrections to masses and mixings in all sectors of the model, some gauge
invariance problems could remain.

\subsection{LanHEP source files}
The \lanhep~ sources  for the MSSM are provided in the
sub-directory MSSM. The  source files have an extension \verb|.src| and contain

\noindent $\bullet$ {\tt var.src} - definition of the independent parameters;

\noindent $\bullet$ {\tt prtcls.src} - list of particles of the
model;

\noindent $\bullet$ {\tt Let.src} - Substitutions to express
field  multiplets in terms of physical particles;

\noindent $\bullet$ {\tt W.src} - superpotential: F-terms and Yukawa interactions; 

\noindent $\bullet$ {\tt DD.src} - scalar supersymmetric potential,  D-terms;

\noindent $\bullet$ {\tt  softsbt.src} -  soft SUSY breaking terms;

\noindent $\bullet$ { \tt fgauge.src} - Gauge fixing terms and Faddeev-Popov ghost for
the gauge group $SU(3)\times SU(2)\times U(1)$;

\noindent $\bullet${\tt ggi.src} - self-interaction of gauge multiplets;
;

\noindent $\bullet${\tt gmi.src} - Interactions of the gauge and matter multiplets;

\noindent $\bullet${\tt higgs4.src} -  effective potential  for the Higgs sector;  

\noindent $\bullet${\tt func.src} - external functions for implementaton of
high order corrections, SLHA interface as well as computation of mass eigenvalues when the
tree-level option is chosen. 

\noindent $\bullet${\tt startup.src} -  main file including the instructions to read  
all other source  files.

To compile the model, launch\\

\verb|lhep startup.src -evl 2|\\

The source files generate three different versions of the MSSM at the electroweak symmetry breaking
scale. The default setting \\

\verb|keys SLHA=On|\\

\noindent
in  \verb|startup.src| compiles the model files in the format needed for an SLHA interface while the
setting
\verb|keys SLHA=Off| will generate the tree level model that requires the matrix diagonalisation
routines for calculating the spectrum. The \\

\verb|keys LambdaTH=On/Off.|\\

\noindent
defines the setting for the computation of the Higgs potential described below.

The running of quark masses is included by defining effecive
masses for quarks as described in section~\ref{sec:qcd}. These masses depend on an external parameter 
$Q$ which specifies the QCD scale relevant for
the process under consideration.

Note that to include the  SLHA model files created in this example into \calchep~ it is necessary to include the
\verb|suspect2.exe| file in the \calchep~ working directory.

\subsection{The Higgs sector}

In the MSSM the Higgs sector receives large loop corrections.
In order to have a realistic tree level model, in particular to have a  light Higgs mass heavier
than the experimental limit, it is necessary to include some higher order corrections to the Higgs sector. 
To do this in a gauge invariant manner, we introduce an effective Lagrangian 
with five independent parameters, $\lambda_1$ to $\lambda_5$.\footnote{The most general Lagrangian needs 7 parameters but
$\lambda_6$ and $\lambda_7$ are in general small
and can be neglected.}  We provide three different methods to include loop corrections in the Higgs sector. 
In the first, the analytical formulae for the one-loop QCD and SUSY-QCD corrections to  the effective Lagrangian~\cite{Carena:1995bx} 
are included in the model file. The masses of the Higgs particles are then computed from 
the effective Lagrangian,  for this one chooses the settings \verb|SLHA=Off|. 
The last two options, which correspond to the setting \verb|SLHA=On|, are available when the higher order corrections
to the Higgs masses and mixing angle are computed by an external program.
In the second method,  the
parameters of the effective Lagrangian are reconstructed from the physical masses, $m_h,m_H,m_{H^+}$ and the Higgs
mixing angle that are provided through the SLHA interface ~\cite{Haber:1997dt}. Here we follow the procedure described in ~\cite{Boudjema:2001ii}.
As there are only four independent
physical parameters, the analytical formulae for the corrections to $\lambda_1$ are used even in
this case, the coefficients $\lambda_2...\lambda_5$ are then extracted from the physical parameters. 
For this option one chooses the setting  \verb|LambdaTH=Off|. This implementation guarantees that the
higher order corrections to the Higgs sector are taken into acount in a gauge invariant
way~\cite{Boudjema:2001ii}. 
Finally a third possibility consists in substituting analytical  formulas for the Higgs couplings~\cite{Carena:1995bx} 
instead of solving equations to extract them from the masses. In this case one must set \verb|LambdaTH=On|.  
Since the spectrum calculators include  in general additional higher-order corrections in the computation of the Higgs masses, 
the masses obtained with the first option differ from the last two, similarly the Higgs potentials can be slightly different in the
last two cases.

\end{document}